# Multichannel sensing platform: galleries of tunable hot spots in ring-disk dielectric dimer with rectangular cross-sections

A.P. Chetverikova[1], N.S. Solodovchenko[1], M. E. Bochkarev[1], K.A. Bronnikov[1], K.B. Samusev[1,2], M.F. Limonov[1,2]

[1]School of Physics and Engineering, ITMO University, St. Petersburg, 191002, Russia
[2]Ioffe Institute, St. Petersburg 194021, Russia

The creation of highly enhanced and localized electromagnetic fields in dielectric structures with low losses, that is, hot spot technology, is an important applied and fundamental problem. However, known examples are reduced to single hot spots generated by low-frequency dipole resonances. Here, a manifold of hot spots were observed experimentally and in calculations in the gap of a ring-inner disk dielectric dimer with a rectangular cross-section. In addition to dipole hot spots, galleries of equidistant hot spots were observed, generated by galleries of disk modes, as well as by galleries of ring modes, each of which begins with a transverse Fabry-Pérot resonance. Therefore, this direction in photonics is called Fabry-Pérot-tronics by analogy with Mie-tronics of spherical particles. The intensity contour of hot spots in the gap has a Fano contour associated with the interference of radiation from narrow longitudinal modes of the ring or disk with broadband scattering on the opposite side of the gap. Controlled transfer of hot spots along the gap and splitting of the hot spot by scanning the exciting frequency in the spectral region of the Fano contour are also implemented. Galleries of tunable bright hot spots provide a new platform for multichannel sensing, SERS and nonlinear applications.

## 1. Introduction

The study of hot spots (HSs) in dielectric resonators is one of the areas of Mie-tonics or dielectric photonics, which began to develop actively after the experimental observation of an induced magnetic response in dielectric particles with Mie resonances [1-5]. Hot spots are areas with high concentration of electromagnetic fields confined in the interparticle gaps. Dielectric photonics is an ideal candidate for the implementation of new platforms in many interdisciplinary areas of research, opening up outstanding prospects in the search for artificial optical magnetic materials. The employment of dielectric nanoparticles with a high refractive index has facilitated the achievement of a remarkable magnetic response overcoming the inherent losses of plasmonic materials. Currently, both the classical Mie resonances, as well as more intriguing interference resonances, such as the Fano resonance [6, 7], bound states in the continuum [8, 9], and anapole mode [10, 11], are being actively studied.

One of the most important tasks of modern photonics is to enhance light-matter interaction by increasing the local electric field strength. Unfortunately, in the case of diffraction-limited dielectric structures, lenses cannot focus light into a spot smaller than $\sim \lambda/2$, and resonators have electromagnetic mode volumes limited to $\sim (\lambda/2)^3$, where $\lambda$ is the wavelength of light in the dielectric. An elegant solution is to create HSs where, due to near-field enhancement, light is trapped and concentrated in a subwavelength gap between two or more dielectric resonant particles located in close proximity [3, 12-14]. A similar field enhancement in the interparticle gap is also observed in clusters of subwavelength particles known as oligomers [15] and in quadrumers of silicon cubes [16]. Recently, the creation of metasurfaces with HSs formed by disk-ring silicon dimers [17] and silicon elliptical nanorods [18] has been reported.

The near-field enhancement of the magnetic component (magnetic HS) and the electric component (electric HS) are being studied experimentally and theoretically [12, 19-21], and both magnetic and electric HSs can be observed in the gap between the same pair of resonators, for example, for a dimer of Si spheres, depending on the excitation geometry [13]. Another interesting feature of HS is that they can arise not only outside, but also inside the dielectric particle [22, 23], moreover, entire hot circles have been demonstrated in a dielectric mesoscale Janus particle [24]. It is believed that despite their modest Q values (typically 10 to 100), metallic resonators can outperform dielectric resonators with much higher Q ($\sim 10^6$) due to their wider bandwidth response [19]. Our work opens a new page in the study of HSs and their practical application, demonstrating the multichannel mode of HSs, which increases the spectral operating range of dielectric structures by orders of magnitude and allows multichannel information processing.

Here, we experimentally and theoretically demonstrate a ring-disk dielectric dimer with multiresonant galleries of equidistant HSs over a wide spectral range while maintaining a high electric field concentration in each HS. Our result is determined by the low-frequency set of photonic eigenmodes of dielectric resonators with rectangular cross-section, such as ring, split ring, and cuboid. The low-frequency photonic spectra of these resonators consist of separate galleries, each of which begin with an intense broad radial or axial Fabry-Perot resonance and continue with an equidistant set of narrow azimuthal resonances, whose quality factor increases exponentially with increasing azimuthal index $m$, [25, 26]. This new direction in photonics we called Fabry-Pérot-tronics by analogy with Mie-tronics of spherical particles. In structures with a rectangular cross-section (ring, split ring, cuboid), two independently controlled Fabry-Pérot resonances give the Fabry-Pérot-tronics an advantage over the Mie-tronics with the radius of the sphere as the only variable geometric parameter.

## 2. Results and Discussion

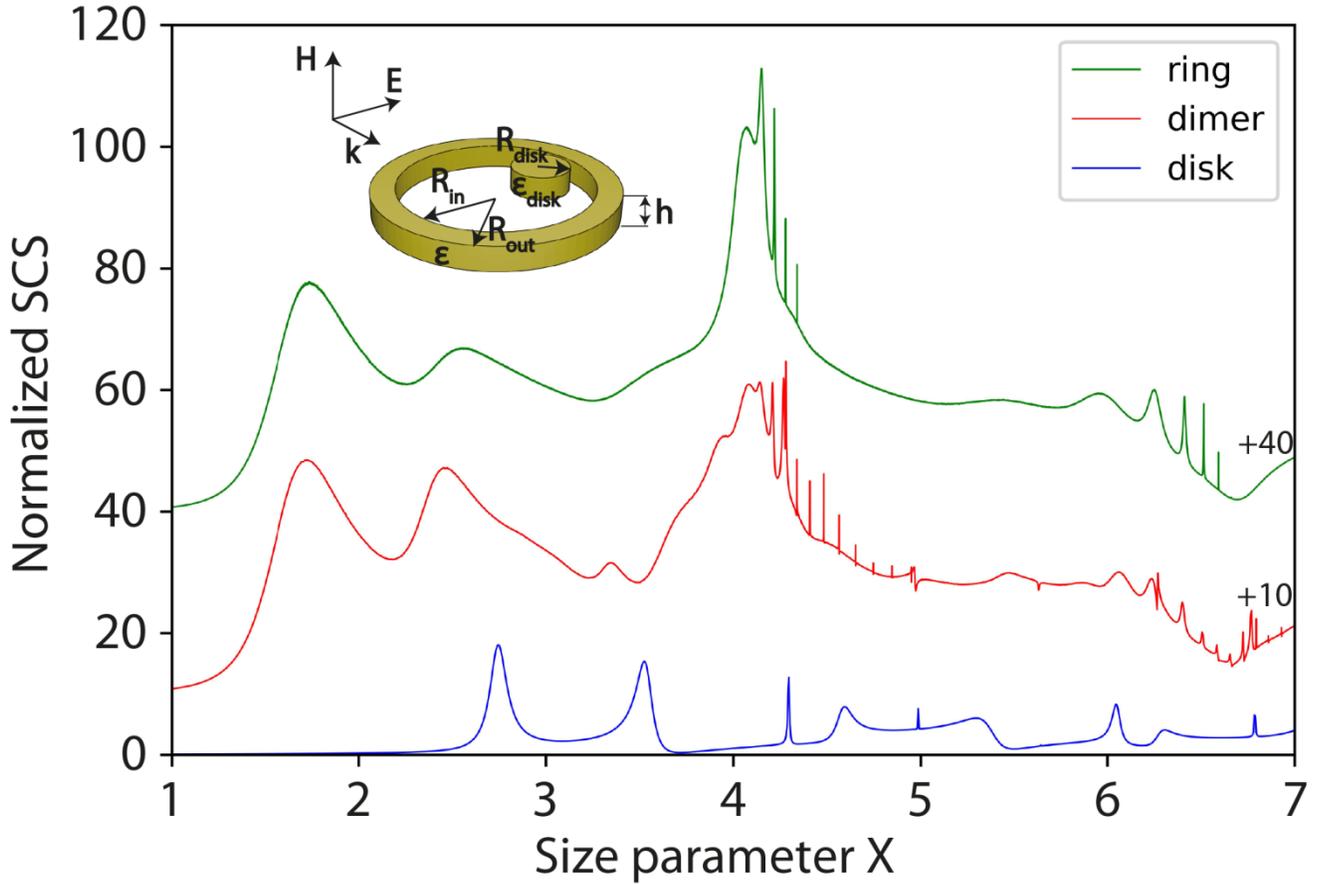

**Figure 1.** Calculated normalized SCS spectra for the dielectric ring (green), dimer (red) and disk (blue) in the range of the size parameter $x=R_{out}\omega/c$ from 1 to 7. Structure parameters: for ring outer radius $R_{out}$ = 56.15 mm, inner radius $R_{in}$ = 45.5 mm, height h = 4 mm, for disk: $R_{disk}$ = 15 mm, $h_{disk}$ = 3.96 mm. The width of the ring-disc gap along the radius of the ring passing through the center of the disk $\Delta=0.01 \cdot R_{in}$ = 0.46 mm. Dielectric constant of the ring $\varepsilon$ = 44.47, dielectric constant of the disk $\varepsilon_{disk}$ = 44.70. TE-polarized incident wave.

**Figure 1** presents the calculated normalized scattering cross section (SCS) of the structures under study, namely, a dielectric ring and a disk of the same height (both structures with rectangular cross sections) and the dimer formed by them, in which the disk is placed inside the ring. The disk has a smaller radius than the inner radius of the ring, so it is possible to realize a dimer with an arbitrarily small gap between the ring and the disk. It has been demonstrated that the low-frequency SCS of a dielectric RR with rectangular cross-sections consists of individual galleries of photonic modes, each of which begins with a broad radial Fabry-Pérot resonance defined by two side walls and continues with a limited set of equidistant longitudinal modes with exponentially increasing quality factors [25]. In addition, Fano resonances [6, 7] were discovered in the scattering spectra of dielectric RRs and strict periodicity of the contours of broad Fabry-Pérot resonances was established according to the Lorentz-Fano-Lorentz-Fano law [25]. In the spectral range shown in Fig. 1, two galleries fit, with the first Fabry-Pérot resonance (radial index $r=1$) has the Lorentzian band shape, the second Fabry-Pérot resonance (radial index $r=2$) has a Fano contour.

Low-frequency SCS of a dielectric disk, shown at the bottom in Fig. 1, has also been studied previously [25]. The new result is the SCS of the ring-disk dimer shown in the middle of Fig. 1 for a small ring-disc gap Δ=0.01·$R_{in}$. In addition, the Supporting Information shows the dependence of the SCS on the position of the disk when it is displaced from the center to the inner wall of the ring (Figure SI-2 in Supporting Information) and the dependence of the ring SCS on the scattering geometry when the direction of the wave vector **k** changes from the ring plane to a direction perpendicular to the ring (Figure SI-3).

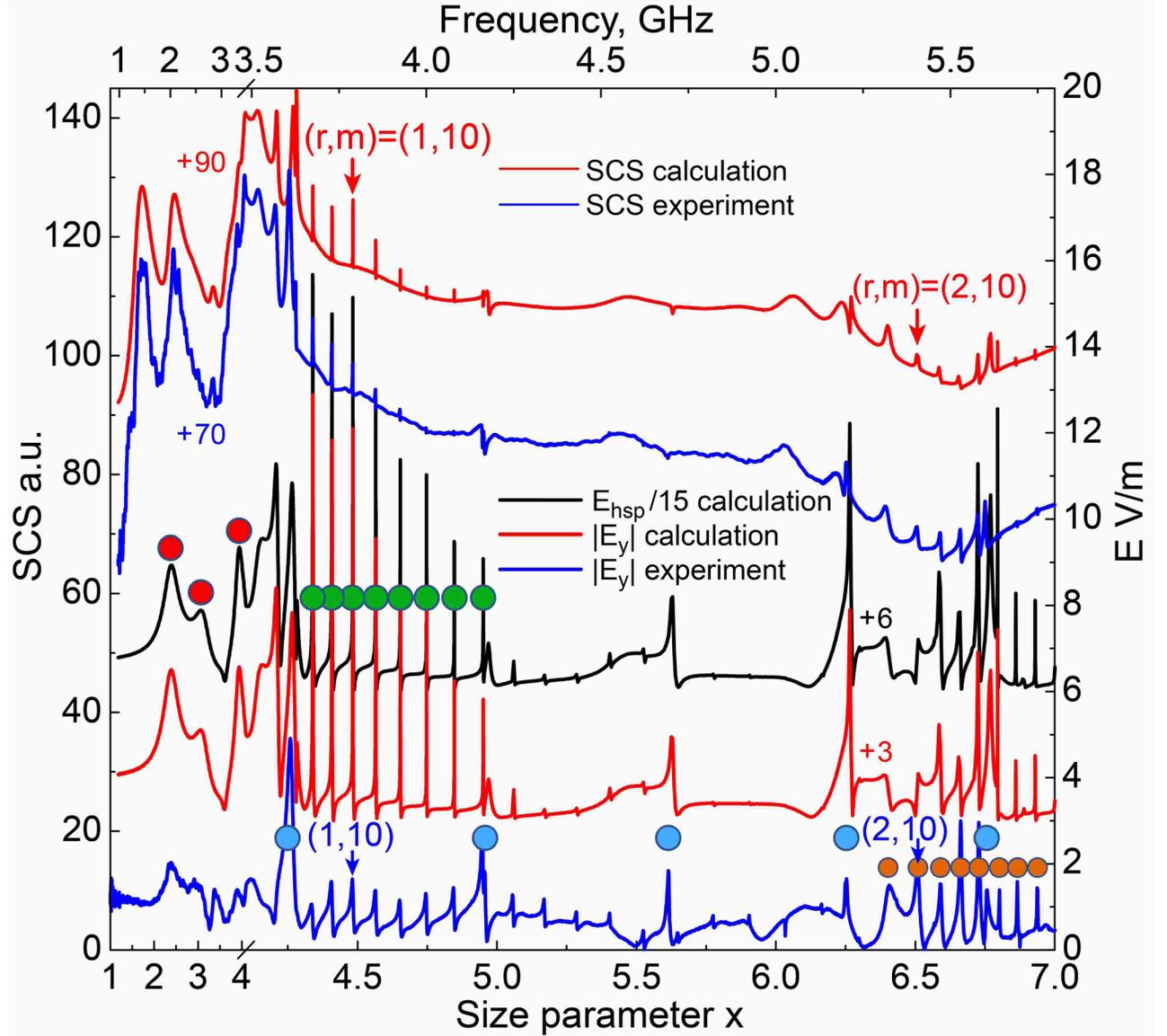

**Figure 2.** The following spectra are shown (from top to bottom): the normalized calculated (red) and experimental (blue) SCS (a.u.) spectra of the dielectric ring-disc dimer; the normalized calculated spectral dependence of the EFS ($E_{hsp}$, Ev/m) at the center of the gap; the calculated (red) and measured (blue) y-component of the EFS ($|E_y|$, Ev/m) at a height of 2 mm above the slit. The circles indicate HSs that are generated by low-frequency modes of the dimer (red), the first (green) and second (orange) photonic galleries of the ring, and the photonic gallery of the disk (blue). For definiteness, modes (1, 10) and (2, 10) of the first and second ring galleries and their corresponding HSs are marked. The width of the ring-disc gap Δ=0.01·$R_{in}$ = 0.46 mm. TE polarization, the size parameter x=$R_{out}ω/c$.

Figure 2 shows the two types of spectra that are the main result of this work. The two upper spectra in the Figure are the calculated and experimental scattering spectra of the dimer in the region of the first and second photonic galleries of the ring. Below the scattering spectra, Figure 2 shows two calculated and one experimental electric field strength (EFS) spectra (Ev/m), demonstrating sharp resonant features corresponding to electrical HSs. The black spectrum shows the calculated spectral dependence of the HSs at the center of the gap between the ring and the disk. The two lower spectra show the calculated (red) and measured (blue) spectral dependences of the electric field strength at a height of 2 mm above the slit, where the near-field sensor was located, measuring the y-component of the electric field $|\mathbf{E}_y|$. The main conclusion that follows from Figure 2 is the ideal coincidence of the frequency position of the HSs with the position of the photonic eigenmodes of the dimer. Another, no less important conclusion is the excellent coincidence of the frequencies of the calculated and experimental resonances. In this case, the difference in the shape of the lines of the calculated and experimental spectrum of $|\mathbf{E}_y|$ is obviously explained by the fact that the sensor installed above the slit collects a signal from an area of 3x3 mm$^2$, and the calculation was carried out for $|\mathbf{E}_y|$ at one point located at a sensor height of 2 mm above the gap opposite the middle of the gap.

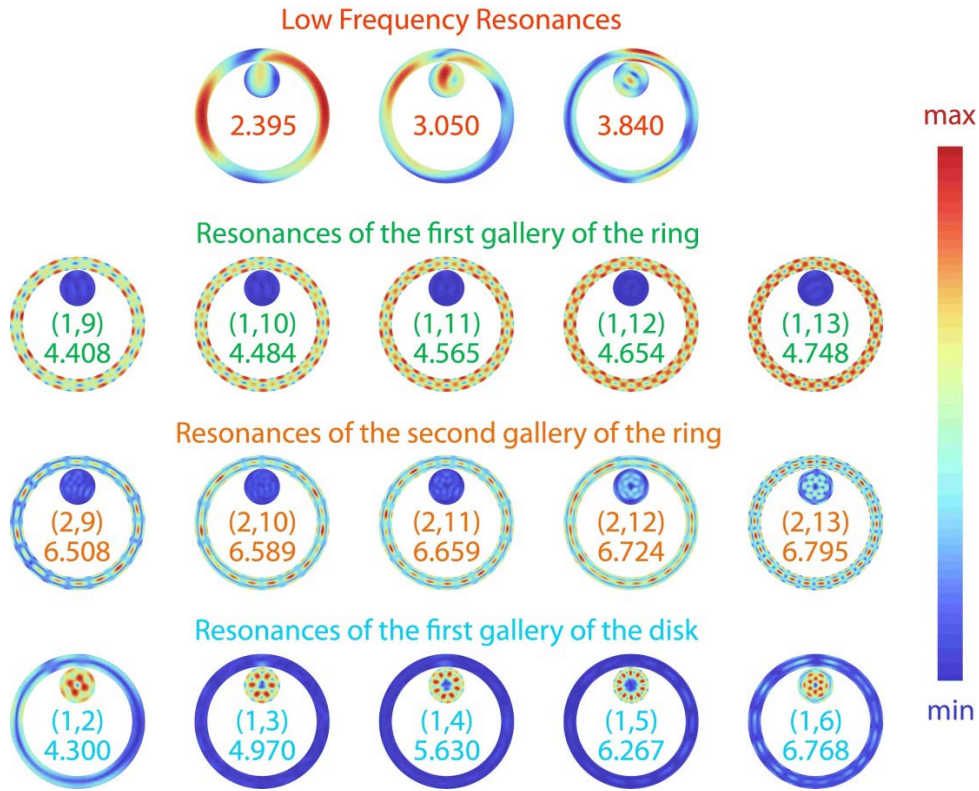

**Figure 3.** Examples of field distribution in the dimer for four types of resonances generating HSs: low-frequency resonances (the size parameters *x* are indicated) and representatives of three galleries, for which the size parameters and radial and azimuthal indices (*r*, *m*) are indicated.

Considering the exact coincidence in line frequencies in the SCS spectra and in the electric field strength spectra, we can confidently interpret the origin of all HSs. For clarity, **Figure 3** shows the field distribution in the ring and disk for low-frequency dimer resonances (three modes in the top row) and for the ring and disk galleries falling within the spectral range under study. In Figure 2, the red circles highlight three low-frequency resonances that have fields resembling dipole modes. It is precisely such points, caused by dipole and octupole modes that were previously observed in calculations and experiments [3, 12-14, 16, 18, 20, 21]. The new result consists of the observation of three galleries of equidistant HSs, which are generated by two galleries of longitudinal modes of the ring and whispering gallery modes of the disk, Figures 2,3. In Figure 2, the HSs of the first and second galleries of the ring are marked with green and orange circles, and the HSs generated by disk modes are marked with blue circles. Note that in the studied spectral range in most cases the modes of the galleries of the ring and the disk do not coincide in frequency, therefore, in Figure 3, either the disk or the ring is dark blue, indicating the absence of resonant excitation. The exceptions are the (2, 12) and (2,13) ring modes and the (1,2) disk mode, when the entire dimer is excited.

There is a significant difference in the profiles of the three types of HSs. Low-frequency quasi-dipole HSs have very broad contours, repeating the low-frequency photonic modes of the dimer with a width of the order of x~0.5 (note the different scales on the abscissa axis before and after x=4). The lines of the HSs of the disk are significantly narrower than the three low-frequency lines, their width is of the order of x~0.05. However, the narrowest lines in the dimer spectra are the longitudinal azimuthal modes of both galleries of the ring $(1,m)$ and $(2,m)$. As noted earlier, the Q factor of the azimuthal modes of an individual dielectric ring without material losses increases exponentially in each gallery with increasing index $m$, exceeding a value of $10^8$ [25]. In our calculations, the Q factor of the HSs in the first ring gallery increased from $10^3$ to $10^4$ with an increase in the $m$ index from 8 to 14; the Q factor of the HSs, measured experimentally above the gap, is about $10^2$.

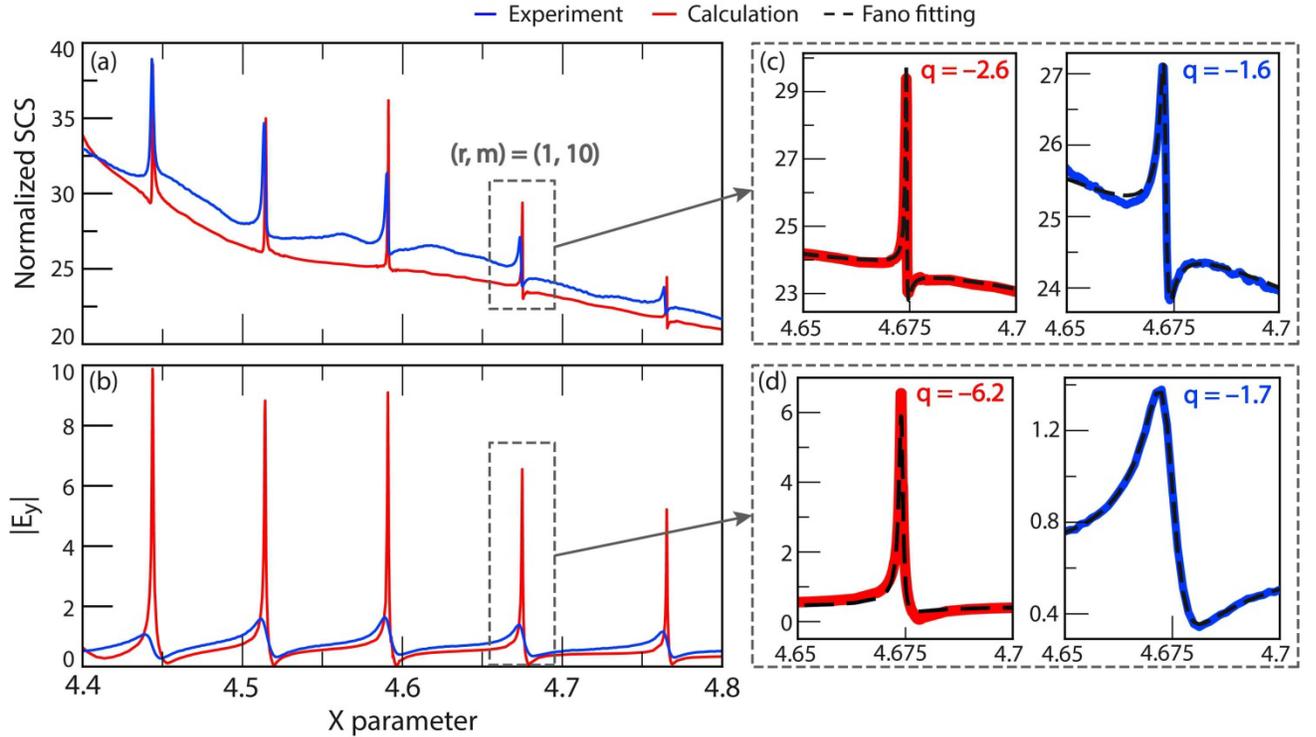

**Figure 4.** (a) Experimentally measured and calculated SCS for the ring-disc dielectric dimer. (b) Experimentally measured and calculated absolute value of electric field component $E_y$ in the middle of the gap 0.46 mm wide. (c) and (d) corresponding Fano fitting for experimental and numerical results for the azimuthal mode with indices $(r, m) = (1,10)$. The size parameter $x=R_{out}\omega/c$.

**Figures 2, 4** clearly demonstrate another important result. All narrow lines, calculated and experimental, in the SCS and HS spectra, have a characteristic asymmetry, more or less pronounced. It is important to note that the scattering spectra demonstrate this asymmetry in the far field (Fig.4a,c), and the hot spot spectra in a gap of $\Delta \sim 0.5$ mm, that is, in the near field for gigahertz waves (note that a frequency of 4 GHz corresponds to a wavelength of 7.5 cm). The lines have asymmetrical contours, characteristic of the Fano resonance [6, 7]. We emphasize that in a dielectric dimer we are not observing a single Fano resonance, but entire galleries of classical Fano contours in the calculated and experimental spectra (Figures 2, 4):

$$\sigma = \sum_i \sigma_i \frac{(\Omega_i + q_i)}{(\Omega_i^2 + 1)}, \tag{1}$$

here $q_i$ is the Fano asymmetry parameter, $\Omega_i = (\omega - \omega_i)/2\Gamma_i$ where $\Gamma_i$ and $\omega_i$ are the half-width and frequency for the i-th resonance, respectively. The Fano contour of HSs in a gap may be due to interference between emission from narrow modes of the ring or disk and broadband scattering on the opposite side of the slit.

It should be noted that studies of HS and SERS in plasmonic structures also demonstrate Fano resonances, but the observed effects are completely different. In the spectra of plasmon structures, only single Fano resonances are observed, which are usually caused by the interference of a broad electric dipole and a less broad quadrupole (or higher order) mode. In contrast to the Fano contours we observed in the form of narrow peaks, in plasmonic structures interference leads to the formation of a pronounced dip (extra-transparency) in the extinction response of the system [3, 21, 27-30]. In the case of plasmonic structures, Fano-type interference is described by a generalization of Fano formula that takes into account material losses [31, 32]:

$$S(\omega) = \frac{\left(\frac{\omega^2 - \omega_a^2}{2W_a \omega_a} + q\right)^2 + b}{\left(\frac{\omega^2 - \omega_a^2}{2W_a \omega_a}\right)^2 + 1}, \tag{2}$$

where $\omega_a$ is the resonance central spectral position, $W_a$ gives an approximation of its spectral width, $q$ is an analogue of the Fano asymmetry parameter, and $b$ is the damping parameter originating from intrinsic losses. If the asymmetry parameter $q$ is equal to zero, the plasmonic resonance is symmetrical and appears as a dip in the spectrum with a minimum value determined by the parameter $b$, which distinguishes the Fano plasmonic resonance from the Fano resonance in lossless dielectrics, in which the contour turns exactly to zero at the frequency corresponding to the condition $q_i + \Omega_i = 0$.

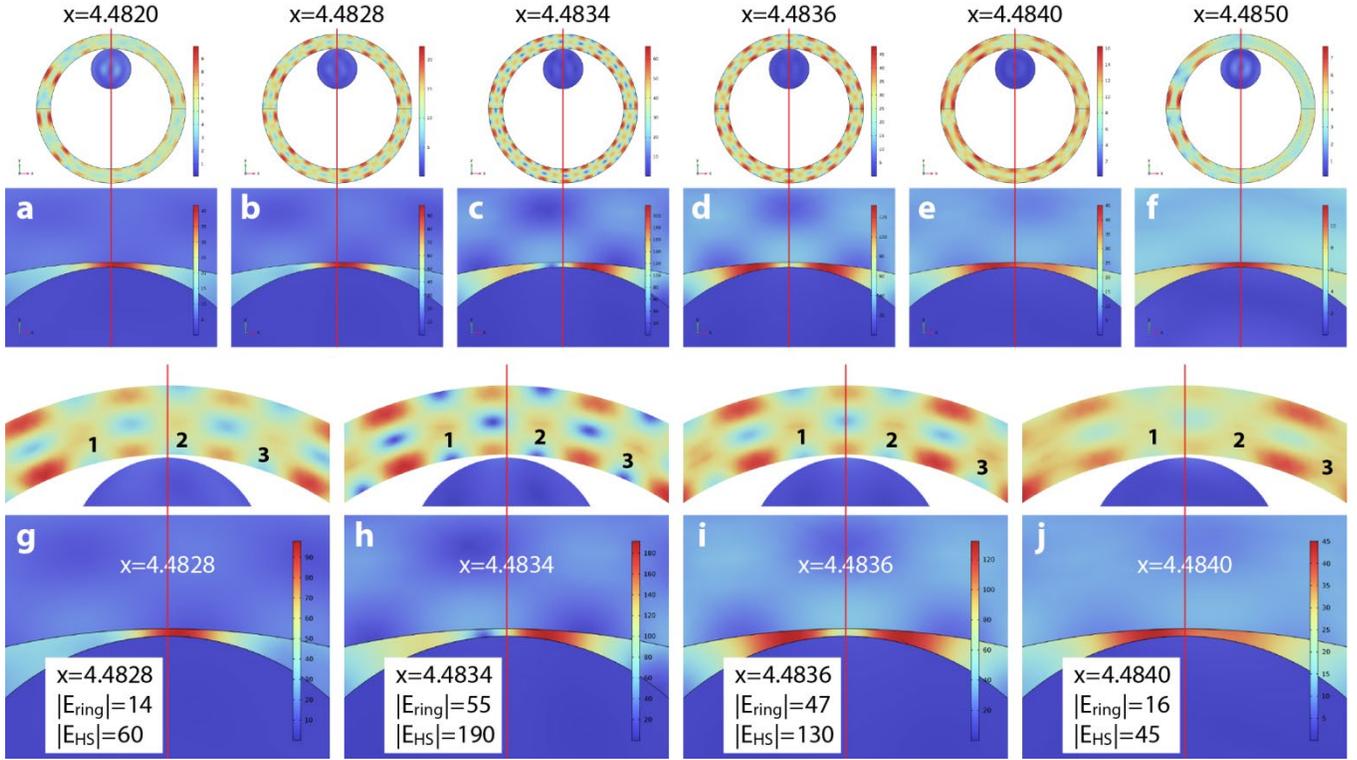

**Figure 5.** HSs engineering and "rotation" of the electromagnetic field in the ring in the Fano resonance regime at a fixed ring-disk gap $\Delta=0.01R_{in}=0.46$ mm. Top row (a-f): Calculated electromagnetic field distribution in the ring near the gap region and field distribution in the gap for six frequencies within the contour of the line (1.10). Bottom row (g-j): Calculated field distribution in the ring and in the gap on an enlarged scale for four frequencies in the Fano resonance region. The red vertical lines are the radius of the ring passing through the center of the disk. The maximum values of the electric field amplitude in the ring $|E_{ring}|$ and in the gap $|E_{HS}|$ are given. The numbers (g-j) indicate three electric field minima, the size parameter $x=R_{out}\omega/c$.

**Figure 5** demonstrates precise control over the locations of HSs in the gap of a dielectric dimer. The key to the controlled transfer of a HS along the gap is Fano interference of electric fields, which in the near field spatially occupies the middle region of the gap. In the spectral intervals before (4.4820≤x≤4.4828) and after (4.4840≤x≤4.4850) the narrow line (1.10), where the field distribution in the ring is not clearly expressed, a weak HS is observed in the narrowest part of the gap, which practically does not shift relative to the symmetry line (the outer panels in the top row of the picture). When scanning the frequency in the region of a narrow Fano resonance (4.4828≤x≤4.4840), two unexpected effects are observed (four panels in the bottom row). The first is related to the behavior of HSs, the second is related to the behavior of the electromagnetic field in the ring. When scanning the frequency in the region of the Fano resonance, the HS is first observed to transfer clockwise along the gap (panel a), then a clear EFS minimum appears (dark blue spot in the slit in panel b), then a split spot is observed (panel c), which eventually merges to one hot spot (panel d).

The transfer of HSs in the gap is directly related to the movement of the electromagnetic field distribution pattern along the ring in the region of the Fano resonance. Note that the HSs in the gap exactly follow the magnetic field maxima at the inner boundary of the ring. Tracing the movement of the three maxima of the magnetic field, marked on the bottom row with numbers 1-3, we observe a displacement of these maxima, inducing the transfer and splitting of HSs in a narrow gap. Thus, the locations and even splitting of HSs in the ring-disk gap can be efficiently controlled by changing the frequency of the external electromagnetic field in the Fano resonance region.

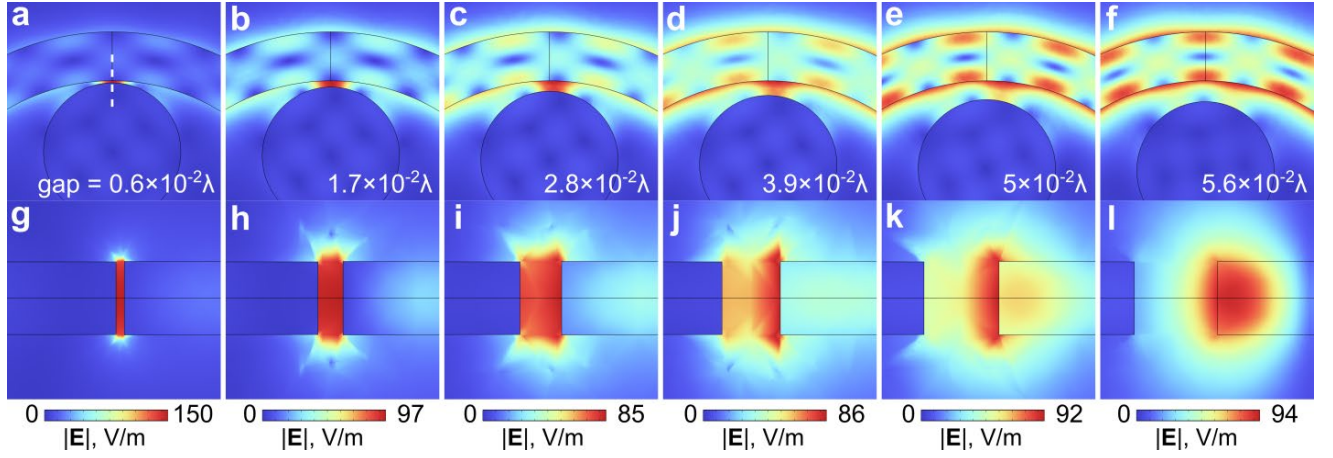

**Figure 6.** Total electric field amplitude |**E**| calculated for the ring mode $(r, m) = (1,10)$ at different gap sizes indicated in pictures; $\lambda$ is the wavelength of incident light. (a-f) Electric field profiles in the XY plane at the half of the dimer's height. (g-l) Electric field profile in the YZ plane along the area indicated as dashed line in (a).

The calculated electric field distributions of the ring mode $(r, m) = (1,10)$ for different gap sizes are presented in **Figure 6**. Here, we denote the gap size in terms of fractions of the incident light wavelength $\lambda$. One can see that for the smallest gap with the size of $0.6 \times 10^{-2}\lambda$ there is a bright single hot spot located at the symmetry axis of the dimer (Figure 6a). The electric field of the hot spot fills the entire gap in the vertical direction (Figure 6g) reaching the maximum amplitude of $|\mathbf{E}| \approx 150$ V/m. Given the incident field strength of 1 V/m, this indicates a 150x electric field enhancement factor. With increasing gap size, the maximum field reduces down to around 100 V/m at the gap size of $1.7 \times 10^{-2}\lambda$ (Figure 6b,h) and to 85 V/m at $2.8 \times 10^{-2}\lambda$ (Figure 6c,i). The field profile of the hot spot becomes more concentrated near the inner wall of the ring. However, for the gap size values of $3.9 \times 10^{-2}\lambda$ (Figure 6d,j) and larger, the maximum electric field in the hot spot area experiences gradual growth and the field in the ring volume becomes comparable to that of the hot spot. Since we analyze one of the ring's eigenmodes, moving the disk away from the ring surface decreases coupling between the ring and disk, which leads to higher Q-factor of the ring mode and, therefore, higher field amplitude inside the ring and in the near-field region. At the gap sizes of $5 \times 10^{-2}\lambda$ (Figure 6e,k) and $5.6 \times 10^{-2}\lambda$ (Figure 6f,l) the hot spot effectively vanishes and transforms into the near-field of the ring. Similarly, the effect of the field pattern rotation around the ring axis is notable in Figure 6c-f. This can be associated with the transition between two modes with the same azimuthal index $m = 10$, but symmetric and antisymmetric field patterns. These modes appear as a result of mode splitting caused by the introduction of the disk, which breaks the axial symmetry of the ring resonator.

## 3. Conclusion

In conclusion, we propose the ring-disk dielectric dimer as a new functional platform for multi-channel information processing, including sensing, nonlinear optics, super-resolution optical imaging.

Equidistant galleries of axial modes of the ring and WGM of the disk generate equidistant galleries of HSs, while the electrical HS appears in a gap opposite the maximum of the magnetic field on the inner side of the ring or the outer side of the disk. A strong enhancement of the electric field at the HS by more than two orders of magnitude has been demonstrated. Precise control over the location and manipulation of HS was demonstrated in two ways - firstly, by scanning the frequency of the exciting wave in the spectral region of the Fano resonance and, secondly, by modulating the geometric size of the gap between the ring and the disk. These results mark a big leap from the well-known reports of single HS in metallic and dielectric resonators of various configurations resulting from electric and magnetic dipole resonances. Demonstrated properties in combination with multi-channel mode are of great importance to realize high-performance filters, sensors, and modulators for prompting applications.

## 4. Experimental Section

*Simulation and Numerical Methods:*

The calculations were performed in the frequency domain using commercial COMSOL software. In order to obtain sufficiently accurate solutions by numerical methods a physics-controlled mesh with the "extremely fine" option was used to capture the geometric details and to resolve the curvature of RRs boundaries. To correctly calculate the scattering cross section (SCS), the resonator was surrounded by a spherical perfectly matched layer and the scattering boundary conditions at its outer boundary were used. The obtained spectra were normalized on the geometric cross section $S = 2R_{out} \cdot h$. The distribution of the electric field was found as the eigenfunctions of the system.

The electric field distributions of the hot spot for varying gap sizes were calculated using COMSOL Multiphysics in the Frequency Domain. A plane wave was incident on the dimer along the X axis with the polarization directed along the Y axis. Perfectly Matched Layer (PML) domains surrounded the modeling area to simulate free space. For each gap size, the field distributions were calculated for the eigenmode of the ring resonator with the azimuthal number $m = 10$ at the frequency corresponding to the maximum of the electric field amplitude in the gap.

*Experimental Methods:*

Far field and near field experiments were all conducted in the microwave chamber. For the far field test of the deflector, as shown in Figure 5e, the sample was fixed in the center of the rotary table. The transmitting horn emitted plane wave which was perpendicular to the surface of the sample. The rotation of the rotary table was controlled and the transmittivity of different angles can be measured with a fixed receiver horn in far field. The transmitting and receiving horns were connected to the vector network analyzer Agilent so that the signals can be detected. For the near field test of the deflector, as shown in Figure 6b, the sample was fixed on the edge of the platform. The transmitting horn was placed on one side of the sample and emitted plane wave which was perpendicular to the surface of the sample. Monopole antenna was used as a receiver to detect the distribution of y-polarization electric field on xoz plane on the other side. The near field test of the lens was analogous.

**Supporting Information**

Supporting Information is available from the Wiley Online Library or from the author.

*Acknowledgement*s

The authors thank Elizaveta Nenasheva (Ceramics Co. Ltd., St. Petersburg) for providing the $(Ca_{0.67}La_{0.33})(Al_{0.33}Ti_{0.67})O_3$ ceramic samples for measurements. The authors are funded by the Russian Science Foundation (project No 23-12-00114).

# Supporting Information

# Multichannel sensing platform: galleries of tunable hot spots in ring-disk dielectric dimer with rectangular cross-sections

A.P. Chetverikova[1], N.S. Solodovchenko[1], M. E. Bochkarev[1], K.A. Bronnikov[1],
K.B. Samusev[1,2], M.F. Limonov[1,2]

[1]School of Physics and Engineering, ITMO University, St. Petersburg, 191002, Russia
[2]Ioffe Institute, St. Petersburg 194021, Russia

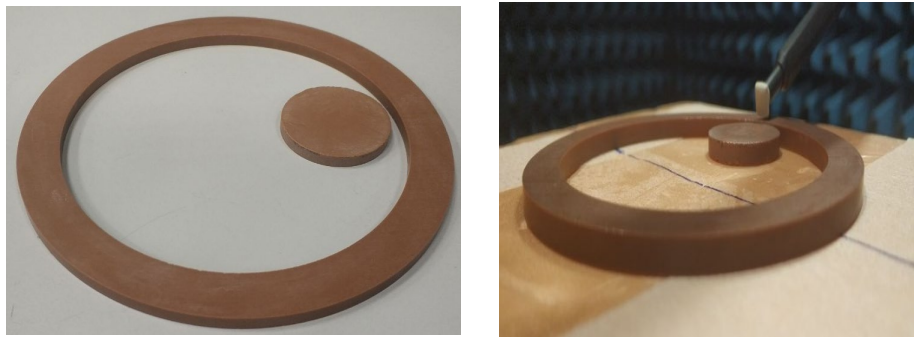

**Figure SI-1.**
Photograph of a ring-disk dielectric dimer (left) and the near-field measurement process in an anechoic chamber (right).

Manufacturer: «CERAMICS» https://ramics.ru

Dielectric parameters of the ceramic composition

| Electrical parameters of ceramics composition | ε | tanδ×10⁴ | ε | tanδ×10⁴ | Q |
|---|---|---|---|---|---|
| | f=1MHz | | f=4.4GHz | | |
| $(Ca_{0,67}La_{0,33})(Al_{0,33}Ti_{0,67})O_3$ | 46,8 | 0,8 | 44.7 | 1,35 | 7410 |

Samples of rings and disks of geometric sizes required for testing and experiments were obtained by hydraulic pressing; 10% solution of polyvinyl alcohol was taken as a binder. The prepared samples were sintered in air within the temperature range of 1540°C (3h) in a chamber electric furnace until zero water absorbance and porosity less than 4% was obtained. To measure the electrical properties at a frequency of 1 MHz, samples of disks with a diameter of 12 mm and a thickness of 2–3 mm were prepared. Measurements of the relative dielectric constant, tanδ, and quality factor Q (1/tanδ) were carried out in the frequency range of ~4.5 GHz on disk samples 14 mm in diameter and 7.5 mm thick by the waveguide dielectric resonator method.

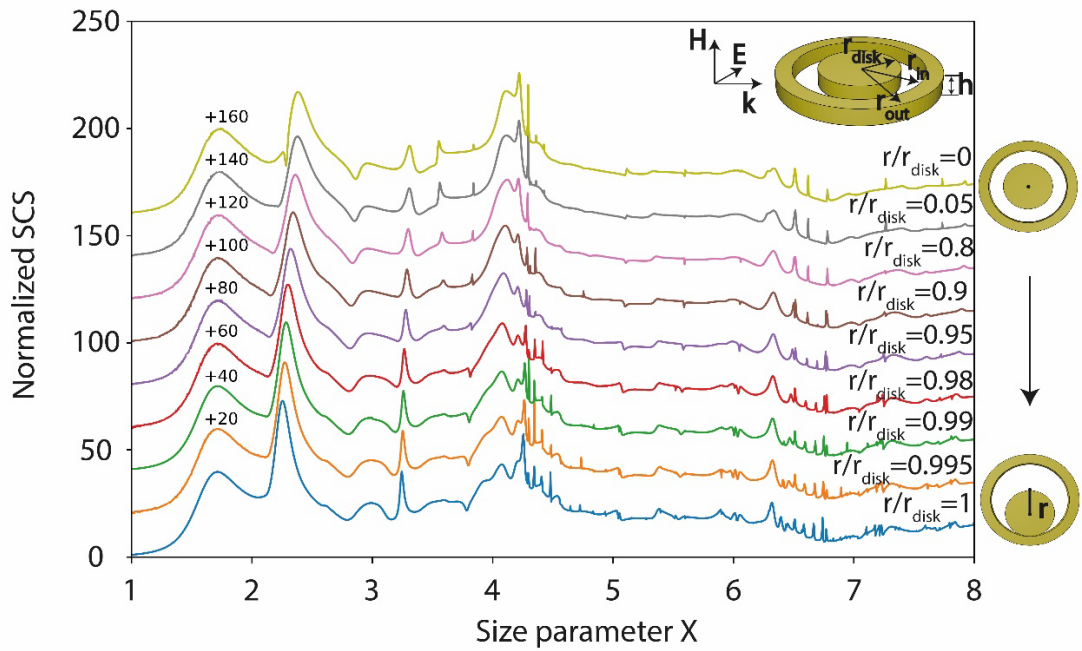

**Figure SI-2.**

Normalized SCS spectra of a dielectric dimer at different disk positions relative to the center of the ring (the position is shown in the figure on the right). Dimer parameters: for the ring, the outer radius $R_{out}$ = 57.5 mm, the inner radius $R_{in}$ = 46.5 mm, height h = 4 mm, for the disk: $R_{disk}$ = 23.25 mm, $h_{disk}$ = 4 mm, $r$ is the distance between the center of the ring and the center of the disk. The dielectric constant of the ring and disk is $\varepsilon$ = 43. TE- polarization. The spectra are shifted vertically by the values shown in the figure on the left.

Total normalized SCS spectra of dielectric RR for various aspect ratio Rout/h indicated to the right of the figure. The spectra are shifted vertically by the values indicated in the figure on the left.

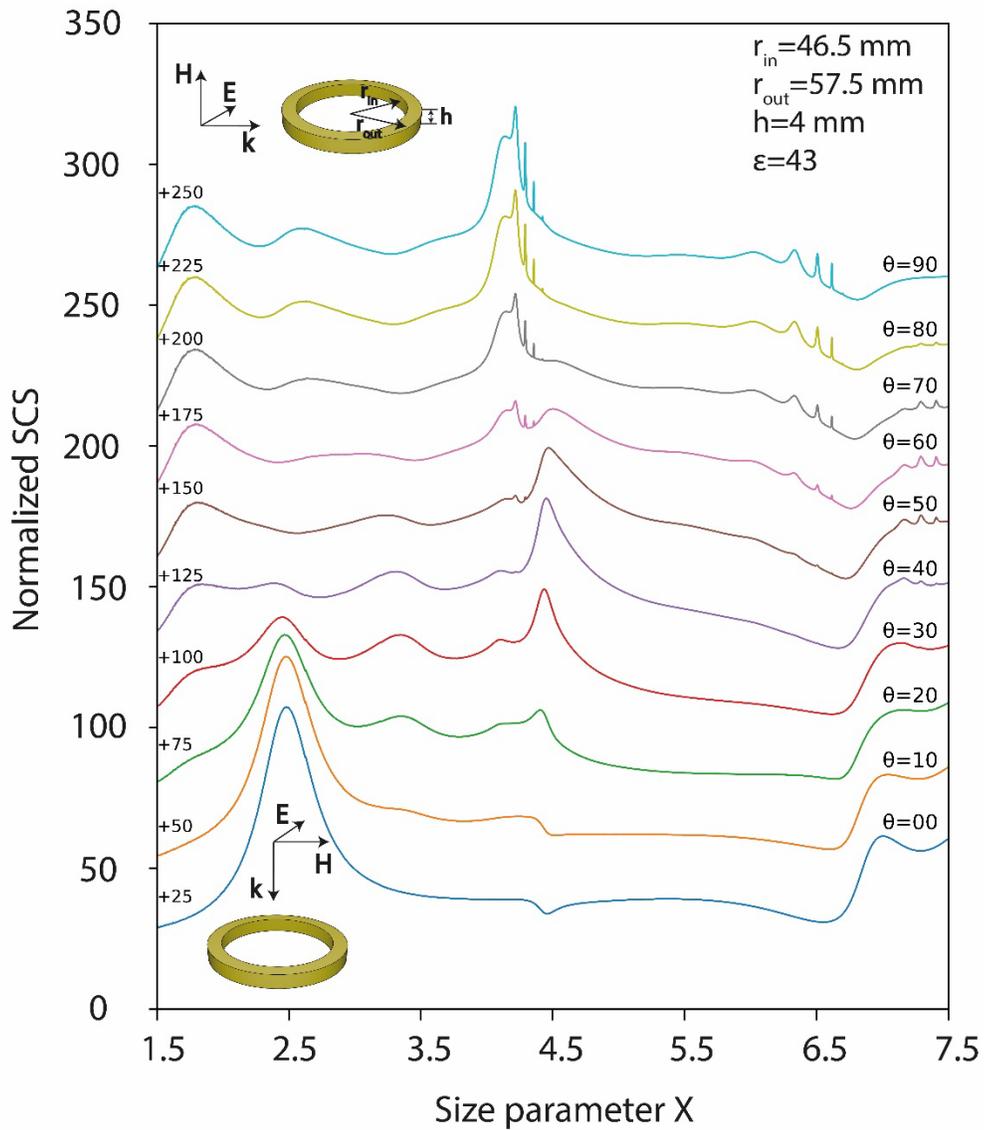

**Figure SI-3.**
Normalized SCS spectra of a dielectric ring at different angles θ of incidence (shown in the figure on the right) of an electromagnetic wave (TE is a polarized wave). Ring parameters outer radius $R_{out}$ = 57.5 mm, inner radius $R_{in}$ = 46.5 mm, height h = 4 mm. The dielectric constant of the ring is ε = 43.
The spectra are shifted vertically by the values shown in the figure on the left.